\documentclass[apj]{emulateapj}
\submitted{ApJ, accepted 15/02/2013}
\usepackage{amsmath}
\usepackage{amsfonts}
\usepackage{amsthm}
\usepackage{amssymb}
\usepackage{graphicx}
\usepackage{mathrsfs}
\usepackage{latexsym}
\usepackage{graphics}
\usepackage{cases}
\usepackage{hyperref}
\usepackage{mathtools}
\usepackage{dsfont}
\usepackage{float}
\usepackage{multirow}
\newcommand{\be}{\begin{equation}}
\newcommand{\ee}{\end{equation}}
\newcommand{\bea}{\begin{eqnarray}}
\newcommand{\eea}{\end{eqnarray}}
\newcommand{\bes}{\begin{equation}\begin{split}}
\newcommand{\ees}{{\end{split}\end{equation}}}
\newcommand{\eq}[1]{eq.~(\ref{#1})}

\renewcommand{\d}{{\rm d}}
\newcommand{\ha}{H{\sc\,i}}
\newcommand{\hm}{H$_2$}
\newcommand{\mha}{M_{\rm H{\sc\,I}}}
\newcommand{\Sha}{\Sigma_{\rm H{\sc\,I}}}
\newcommand{\hh}{h_{70}}
\newcommand{\msun}{{\rm M}_{\odot}}

\newcommand{\hhmsun}{\hh^{-1}\msun}
\newcommand{\sax}{S$^3$-SAX}

\newcommand{\erf}{\rm erf}
\newcommand{\e}{\rm E}
\renewcommand{\sp}{S_{\rm p}}
\newcommand{\sint}{S_{\rm int}}
\newcommand{\vc}{V_{\rm c}}
\newcommand{\w}{W_{50}}
\newcommand{\lcdm}{$\Lambda$CDM}
\newcommand{\kms}{{\rm km~s^{-1}}}
\newcommand{\ms}{M_{\rm stars}}

\newcommand{\jy}{{\rm Jy}}
\newcommand{\hmpc}{\,h^{-1}{\rm Mpc}}
\newcommand{\mdm}{m_{\rm DM}}
\newcommand{\kevc}{{\rm~keV}\,c^{-2}}

\begin{document}

\title{Confronting Cold Dark Matter Predictions with Observed Galaxy Rotations}

\author{Danail Obreschkow$^1$, Xiangcheng Ma$^2$, Martin Meyer$^{1,3}$, Chris Power$^{1,3}$, Martin Zwaan$^4$, Lister Staveley-Smith$^{1,3}$, Michael J.~Drinkwater$^5$}
\affiliation{$^1$International Centre for Radio Astronomy Research (ICRAR), M468, University of Western Australia, 35 Stirling Hwy, Crawley, WA 6009, Australia}
\affiliation{$^2$The University of Sciences and Technology of China, Centre for Astrophysics, Hefei, Anhui 230026, China}
\affiliation{$^3$ARC Centre of Excellence for All-sky Astrophysics (CAASTRO)}
\affiliation{$^4$European Southern Observatory, Karl-Schwarzschild-Str. 2, 85748 Garching b. M\"{u}nchen, Germany}
\affiliation{$^5$School of Mathematics and Physics, The University of Queensland, Brisbane, QLD 4072, Australia\\}

%\pacs{47.55.dp,47.55.dd,43.25.Yw}

%\date{\today}

\begin{abstract}
The rich statistics of galaxy rotations as captured by the velocity function (VF) provides invaluable constraints on galactic baryon physics and the nature of dark matter (DM). However, the comparison of observed galaxy rotations against cosmological models is prone to subtle caveats that can easily lead to misinterpretations. Our analysis reveals full statistical consistency between $\sim\!5000$ galaxy rotations, observed in line-of-sight projection, and predictions based on the standard cosmological model (\lcdm) at the mass-resolution of the Millennium simulation (\ha~line-based circular velocities above $\sim\!50~\kms$). Explicitly, the \ha~linewidths in the \ha~Parkes All Sky Survey (HIPASS) are found consistent with those in \sax, a post-processed semi-analytic model for the Millennium simulation. Previously found anomalies in the VF can be plausibly attributed to (1) the mass-limit of the Millennium simulation, (2) confused sources in HIPASS, (3) inaccurate inclination measurements for optically faint sources, and (4) the non-detectability of gas-poor early-type galaxies. These issues can be bypassed by comparing observations and models using linewidth source counts rather than VFs. We investigate if and how well such source counts can constrain the temperature of DM.
\end{abstract}

\maketitle

%%%%%%%%%%%%%%%%%%%%%%%%%%%%%%%%%%%%%%%%%%%%%%%%%%%%%%%%%%%%%%%%%%%%%%%%%%%%%%%%%%%%%%%%%%%%%%%%%%%%%%%%%%%%%%%%%%%%%%%%%%%%%%%%%%

\section{Introduction}\label{section_introduction}

Mass and angular momentum are crucial galaxy properties, since their global conservation laws constrain the history and future of galaxy evolution \citep{Bullock2001,Bullock2001b}. Moreover, measurements of mass and angular momentum uncover hidden dark matter and potentially constrain its nature \citep{Zavala2009,Obreschkow2013a}. In recent decades, the mass statistics have been studied in detail via the mass function (MF, \citealp{Li2009}), the luminosity function (LF, \citealp{Loveday2012}), and the auto-correlation function \citep{Blake2011a}. By contrast, angular momentum remains a side-topic, normally addressed indirectly via the Tully-Fisher relation (TFR, \citealp{McGaugh2012}) or used as a means of recovering the mass distribution in individual galaxies \citep{DeBlok2008}. Spatial statistics of angular momentum and the related circular velocity function (VF, \citealp{Gonzalez2000,Desai2004,Zwaan2010,Papastergis2011}) remain relatively unexplored. This is despite the fact that the VF offers a tremendous potential with regard to comparing LFs obtained in different wave-bands (\citeauthor{Gonzalez2000}), measuring various mechanisms of feedback in the evolution of galaxies \citep{Sawala2012}, and constraining the temperature of dark matter \citep{Zavala2009}.

\begin{figure}[b]
	\includegraphics[width=\columnwidth]{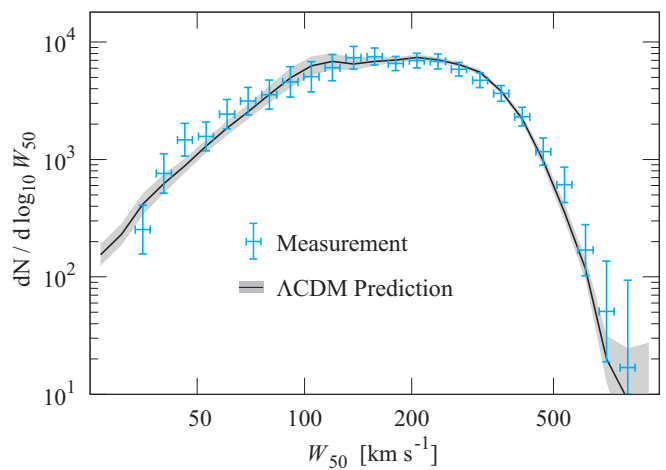}
	\caption{(Color online) Counts of \ha~linewidths in the largest equivalent subsamples of the HIPASS survey and the \lcdm-based \sax-model. This plot shows that measured and simulated linewidths agree on the completeness domain of $\mha\gtrsim10^8~\msun$ and $\vc\gtrsim50~\kms$. Error bars represent 67\%-measurement uncertainties associated with actual measurement noise, cosmic variance, and completeness uncertainties. Similarly, the grey shading depicts 67\%-confidence intervals for the model, associated with cosmic variance. This figure is a simplified version of Fig.~\ref{fig_overview}b.}
	\label{fig_essential}
\end{figure}

In fact, measuring a galaxy's rotational velocity is challenging, since it requires both a measurement of the galaxy inclination, typically drawn from a spatially resolved optical image, as well as a measurement of the line-of-sight rotational velocity, typically obtained from the Doppler-broadening of the 21~cm emission line of neutral hydrogen (\ha). Today, only two \ha\ surveys offer reasonably large samples to construct VFs, the \ha~Parkes All Sky Survey (HIPASS, \citealp{Barnes2001}) and the ongoing Arecibo Legacy Fast ALFA (ALFALFA, \citealp{Giovanelli2005b,Giovanelli2005}). They are the largest surveys by the cosmic volume and by the number of galaxies, respectively. The VFs derived from HIPASS \citep{Zwaan2010} and the 40\%-release of ALFALFA \citep{Papastergis2011} were both compared against theoretical models, including predictions by the \sax-model \citep{Obreschkow2009b}, the only current model of frequency-resolved \ha-emission lines in a cosmological simulation. These comparisons uncovered statistically significant differences, some of which could be attributed to gas-poor massive early-type galaxies (\citeauthor{Zwaan2010}), but the physical implications remained unclear. Differences in the faint-end of the velocity function (Fig.~9 in \citeauthor{Papastergis2011}), near the resolution limit of the \sax-model, seemed to hint a possible breakdown of the current cosmological model. In a new attempt to understand and exploit these differences, we successively found them to be subtle artifacts of the comparison itself, hence motivating a more detailed analysis.

This paper presents a revised comparison between the \ha~line profiles in HIPASS and \sax. We deliberately focus on HIPASS, while reserving a similar analysis of the ongoing ALFALFA survey for the future, because HIPASS already has optical inclinations available, exhibits a detailed completeness function, and contains less cosmic variance than 40\%-ALFALFA in terms of the redshift-distribution of the galaxies (see Fig.~4a by \citealp{Martin2010} versus Fig.~2 bottom by \citealp{Zwaan2005}). The HIPASS data is compared against the \sax-model in various ways. A key result, worth highlighting early, is the full consistency between the 50-percentile \ha~linewidth $\w$ in HIPASS and \sax, as illustrated by the counts in Fig.~\ref{fig_essential}. In this work we compare both apparent \ha\ linewidths and inclination-corrected circular velocities using source counts, as well as space density functions. The different aspects uncovered by these functions are discussed in detail, as well as their reliability as statistical estimators. Based on the results, we finally conjecture that linewidth counts might be a useful tool for measuring the temperature of dark matter, and discuss how well HIPASS can, in principle, constrain this temperature.

The manuscript is organized as follows. Section \ref{section_data} first explains the observed dataset (HIPASS with optical imaging) and its simulated counterpart (\sax). Five statistically independent simulations are generated specifically to assess the effects of cosmic variance. The observed and simulated datasets are then truncated to congruent subsamples suitable for their comparison. This comparison is presented in detail in Section \ref{section_comparison}. In Section \ref{section_discussion}, the consistency between HIPASS and \sax\ is interpreted and discussed with respect to the TFR and alternative models of dark matter. Section \ref{section_conclusion} summarizes the results in a list of key messages.\vspace{1mm}

\section{Data description}\label{section_data}

\subsection{Observed data: HIPASS}\label{subsection_observation}

HIPASS is a blind search for \ha~emission at declinations ${\rm dec}<+25^\circ$ in the velocity range $-1,280~\kms<cz<12,700~\kms$, where $c$ is the speed of light and $z$ is the redshift. This survey resulted in 5317 identified galaxies, gathered in two catalogs: the HIPASS galaxy catalogue (HICAT, \citealp{Meyer2004,Zwaan2004}) containing 4315 sources with ${\rm dec}<+2^\circ$, and its northern extension (NHICAT, \citealp{Wong2006}) containing 1002 sources with $+2^\circ<{\rm dec}<+25^\circ$. The \ha~lines of these 5317 sources have been parameterized in various ways. In this work, we will use the luminosity distance $D_{\rm L}$, given in Mpc, the velocity-integrated line flux $\sint$, given in $\jy~\kms$, the corresponding \ha~mass $\mha=2.36\cdot10^5\sint D_{\rm L}^2(1+z)^{-1}$, given in $\msun$, the peak-flux density $\sp$, given in $\rm mJy$, and the linewidth $\w$ (``$\w^{\max}$'' in HICAT), given in $\kms$ and measured at 50\% of the peak flux density. HIPASS uses a channel width of $13.2~\kms$, but parameterization was carried out after two stages of smoothing (Tukey and Hanning), resulting in a full-width-half-max resolution of $26.4~\kms$ for $\w$.

\cite{Doyle2005} presented optical counterparts for HICAT, identified in the $b_J$-band plates of the SuperCOSMOS Sky Survey \citep{Hambly2001}. To each of these galaxies they fitted an ellipse to measure the semi-major axis $a$, the semi-minor axis $b$, and the position angle. There are 3618 sources in HOPCAT with identified values $a$ and $b$. From those values the galaxy inclinations $i$ can be estimated using the spheroid assumption,
\be\label{eq_i}
	\cos^2 i = \frac{q^2-q_0^2}{1-q_0^2},
\ee
where $q\equiv b/a$ and $q_0$ denotes the intrinsic axis ratio, here taken to be $q_0=0.2$ to remain consistent with \cite{Zwaan2010}. As in the latter work, we here \textit{define} the circular velocity $\vc$ of a galaxy as
\be\label{eq_vc}
	\vc \equiv \frac{\w}{2\sin i},
\ee
although the actual asymptotic rotational velocity may slightly differ from $\vc$.

\subsection{Simulated data: \sax}\label{subsection_simulation}

%\begin{figure*}
%	\includegraphics[width=\textwidth]{fig_simulation}
%	\caption{(Color online) Flow-diagram of the simulation as explained in Section \ref{subsection_simulation}. This work focusses particularly on steps (c) and (d).}
%	\label{fig_simulation}
%\end{figure*}

This section summarizes \sax\ \citep{Obreschkow2009b}, the first cosmological model of resolved \ha-emission lines of galaxies.

\sax\ builds on model-galaxies generated by a semi-analytic model \citep[SAM,][]{DeLucia2007}. The latter relies on the Millennium simulation \citep{Springel2005} that tackles the evolution of cold dark matter (CDM) in a comoving box measuring $(500\hmpc)^3$, where $h$ is defined via the local Hubble constant $H_0\equiv100\,h\rm\,km\,s^{-1}\,Mpc^{-1}$. This simulation uses the standard cosmological model (\lcdm) with parameters $h=0.73$, $\Omega_{\rm m}=0.25$, $\Omega_{\rm b}=0.045$, $\Omega_\Lambda=0.75$, and $\sigma_8=0.9$. From this simulation CDM halos and their merging histories are extracted. The SAM then assigns galaxies to the centers of these halos using a cooling model and evolves global galaxy properties, such as stellar mass, gas mass, and morphology according to physical rules allowing for feedback from black holes and supernovae. The free parameters in this SAM were tuned to the locally observed color-magnitude distribution, but there is no explicit fit to galaxy sizes, rotations, and gas properties. The number of model-galaxies at a time of $13.7\cdot10^9\rm~yr$, i.~e., today, is to about $3\cdot10^7$. Although the cosmological parameters of the Millennium simulation are slightly inconsistent with the newest estimates \citep{Komatsu2011}, the present-day galaxy properties remain nearly unaffected according to calculations by \cite{Guo2013}.

\begin{figure}[t]
	\includegraphics[width=\columnwidth]{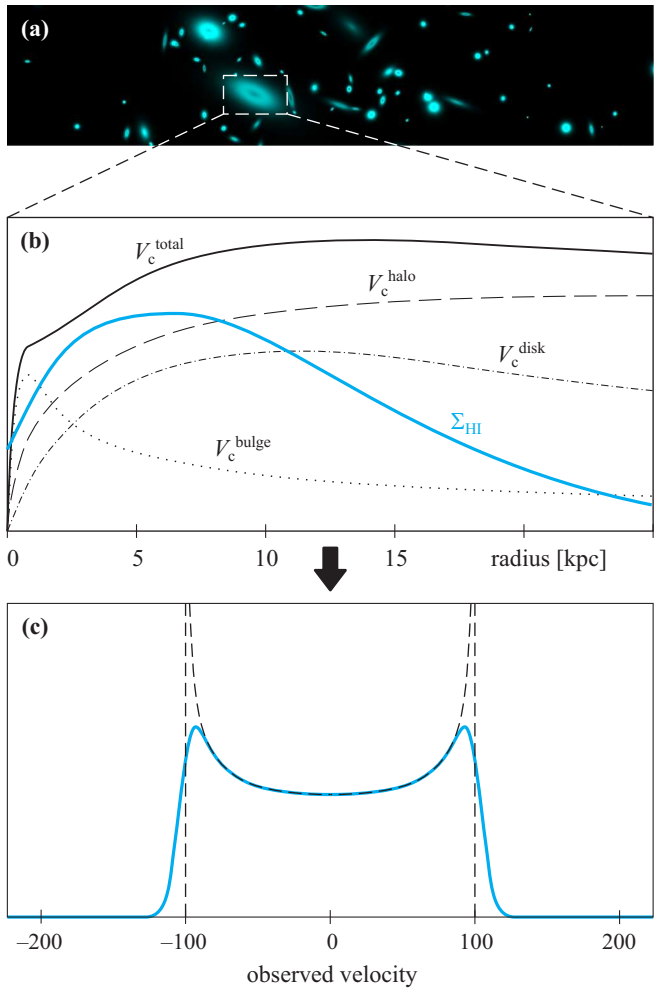}
	\caption{(Color online) Illustration of our method to model \ha~emission lines. For each model-galaxy, we compute an axially symmetric \ha~density profile $\Sha(r)$ and a circular velocity profile $\vc^{\rm total}(r)$ based on mass distribution in the halo, disk, and bulge. Convolving $\Sha(r)$ with $\vc^{\rm total}(r)$ yields an ideal edge-on emission line (panel c, dashed line), which is smoothed with a Gaussian Kernel of $\sigma=8~\kms$ to account for the velocity dispersion typical for the local universe (panel c, solid line).}
	\label{fig_s3sax}
\end{figure}

Given the evolving model-galaxies of the SAM, \cite{Obreschkow2009b} assigned refined cold gas properties to each galaxy. Their method, sketched out in Fig.~\ref{fig_s3sax}, can be summarized as follows. The scale radius of galactic disks is calculated directly from the spin of the dark matter halo. To do so, a variable ratio $\xi\in[0.5-1]$ between the specific angular momentum of baryons and dark matter was adopted. This ratio is a function of the Hubble-type and the stellar mass, adjusted such that the resulting disk scale radii optimally reproduce those of the real galaxies in The HI Nearby Galaxy Survey (THINGS, \citealp{Walter2008}). Given the disk scale radius and the total cold gas mass from the SAM, the radial \ha~surface density $\Sha(r)$ is calculated using a pressure-based model for the ratio between molecular (\hm) and atomic (\ha) hydrogen \citep{Obreschkow2009a}, derived from the THINGS sample \citep{Leroy2008}. In parallel, circular velocity profiles $\vc^{\rm total}(r)=({\vc^{\rm halo}}^2(r)+{\vc^{\rm disk}}^2(r)+{\vc^{\rm bulge}}^2(r))^{1/2}$ are calculated from the circular velocities implied by the gravitational potentials of the dark matter halo, the galactic disk, and the central bulge, respectively. Convolving $\vc^{\rm total}(r)$ with $\Sha(r)$ then results in a model for the frequency-resolved \ha~emission line (dashed line in Fig.~\ref{fig_s3sax}c), which, when convolved with a Gaussian Kernel for dispersion, turns into a smooth profile (solid line).

\begin{figure}
	\includegraphics[width=\columnwidth]{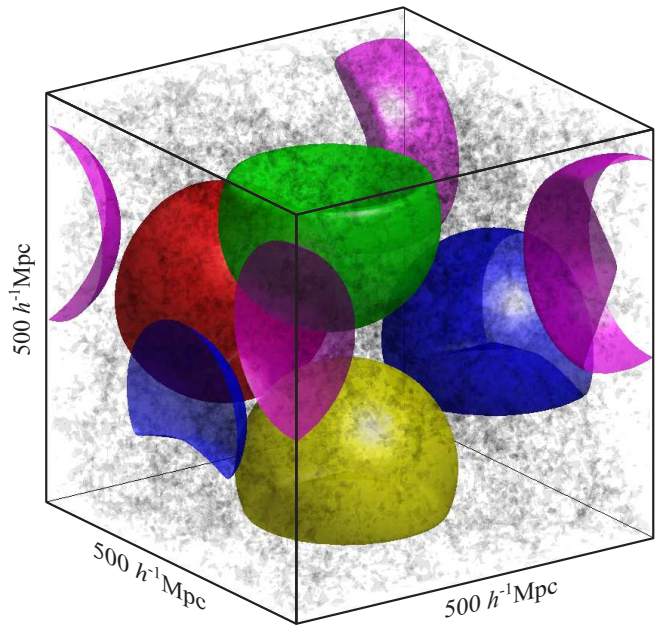}
	\caption{(Color online) The five colors represent our five virtual HIPASS volumes fitted inside the cubic box of the Millennium simulation. The box obeys periodic boundary conditions, such that the truncated volumes (blue and purple) are in fact simply connected. The five HIPASS volumes do not overlap and can hence be used to estimate the magnitude of cosmic variance.}
	\label{fig_millennium}
\end{figure}

Departing from the cubic box of the Millennium simulation populated with model-galaxies with resolved \ha~emission lines, \cite{Obreschkow2009f} produced a sky-model with apparent extra-galactic \ha~emission as seen by a fixed observer. To do so, the Cartesian coordinates $(x,y,z)$ of the model-galaxies were mapped onto apparent positions, i.e., right-ascension (RA), declination (dec), and redshift $z$, using the method of \cite{Blaizot2005}. This method explicitly accounts for the fact that galaxies more distant from the observer are seen at an earlier stage in their cosmic evolution. Along with this mapping, the intrinsic \ha~luminosities are transformed into observable fluxes. Moreover, the \ha~emission line of each galaxy is corrected for the inclination of the galaxy, respecting, however, the isotropy of turbulent/thermal dispersion. The linewidth $\w$ at the 50\%-level of the peak flux density is then measured from the apparent \ha~line of the inclined model-galaxy. To allow a clean comparison with observations, we then calculate the circular velocity $\vc$ of a model-galaxy via eq.~(\ref{eq_vc}). This is an important step, since $\vc$ can differ from the asymptotic value of $\vc^{\rm total}(r\rightarrow\infty)$ by up to $30\%$ for some galaxies with relatively compact \ha~distributions.

For the purpose of this paper we realized five different virtual skies by placing the cosmic volume probed by HIPASS five times inside the simulation box of the Millennium simulation as shown in Fig.~\ref{fig_millennium}. There is no overlap between these five sub-volumes, making them (almost) statistically independent. These five virtual sky volumes will be used to quantify the effects of cosmic variance, that is the random effects of the locally inhomogeneous large scale structure.

\begin{figure}
	\includegraphics[width=\columnwidth]{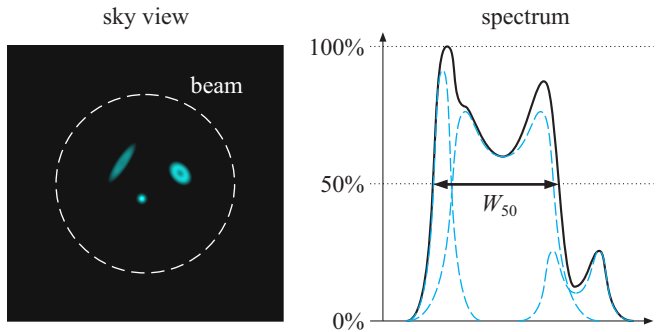}
	\caption{(Color online) Simulated galaxies with centers separated by less than the FWHM of the HIPASS beam of $15.5'$ (left panel) and with \ha~emission lines overlapping in frequency at the 20-percentile level (dashed lines, right panel) are regarded as confused. For comparison with HIPASS, such confused sources are considered as a single sources with an emission line (solid line) obtained by co-adding the individual constituents.}
	\label{fig_confusion}
\end{figure}

In the aim of comparing the \sax-model against HIPASS it is crucial to note that the gridded beam of the HIPASS data measures 15.5' at full-width-half-max. Using the \sax~sky-model, we find that this limited spatial resolution implies a non-negligible probability for two or more \ha~disks to be confused, i.e., to fall inside the same beam and simultaneously overlap in frequency. This confusion must hence be accounted for when comparing observations against simulations. We do so by merging simulated galaxies, whose centroids are separated less than 15.5' and whose \ha\ lines overlap in frequency. The common \ha\ mass is then taken as the sum of the individual components and $\w$ is measured from the combined line as shown in Fig.~\ref{fig_confusion}. We further define the circular velocity $\vc$ of the merged object as the \ha~mass-weighted average of the circular velocities of the components. This procedure reduces the number of simulated sources by about 2\%.

\subsection{Sample selection}

Let us now construct subsamples of sources in HIPASS and \sax\ using identical selection criteria. Two types of samples will be considered depending on whether optical counterparts are required for HIPASS sources. These counterparts are needed when considering circular velocities $\vc$, since those require estimates of the galaxy inclinations. However, masses $\mha$ and linewidths $\w$ do not require optical data. We shall call the larger galaxy sample, in which optical counterparts are irrelevant, the ``reference sample''. A subsample of this reference sample, in which all galaxies have optical inclinations and thus estimates of $\vc$, is then called the ``$\vc$-sample''. The precise selection criteria of these two samples are listed in Tab.~\ref{tab_selection_criteria} and explained in the following.

\subsubsection{Reference sample}

\textit{Volume truncation:} Since the volume of \sax\ exceeds that of HIPASS, the former must be truncated to the field-of-view (FOV) and redshift range of HIPASS, i.e.,~$\rm dec\leq+25^\circ$ and $cz\leq12,700\rm~km\,s^{-1}$. These criteria truncate \sax\ to sub-volumes matching the colored regions in Fig.~\ref{fig_millennium}.

\textit{Mass limit:} The limiting \ha~mass, above which \sax\ can be considered complete, is about $10^8\,\hhmsun$ (\citealp{Obreschkow2009b}), where $\hh\equiv h/0.7$, i.e.,~$\hh=1$ if $H_0=70\rm\,km\,s^{-1}\,Mpc^{-1}$, as consistent with current observations \citep{Jarosik2011}. The simulated \ha~MF drops rapidly below this limiting mass due to the limited mass resolution of the Millennium simulation. We must therefore limit the comparison between simulated and observed sources to the mass range $\mha\geq10^8\,\hhmsun$. This criterion removes 71 nearby galaxies from HIPASS, that is about $1.3\%$ of the 5317 identified sources.

\textit{Limiting linewidth:} HIPASS does not resolve sources with linewidths smaller than $25~\kms$ (see Section 3.1\ in \citealp{Meyer2004}). For correctness we therefore apply the selection criterion $\w\geq25~\kms$ to \sax, although this only reduces the number of simulated sources by about $0.1\%$.

\textit{Completeness limit:} In HIPASS, real sources are detected with a probability approximated by
\be\label{eq_completeness}
	C(\sp,\sint)	= \e\big[p_1(\sp-p_2)\big]\e\big[p_3(\sint-p_4)\big],
\ee
where $\e(x)\equiv\max\{0,\erf(x)\}$. The parameters are $\mathbf{p}=(0.036,19,0.36,1.1)$ for $\rm dec\leq+2^\circ$ \citep{Zwaan2004} and $\mathbf{p}=(0.02,5,0.14,1)$ for $\rm dec>+2^\circ$ \citep{Wong2006}. The same completeness function must be applied to \sax. This is done by drawing a random number $R$ uniformly from the interval $[0,1]$ for every simulated galaxy, and retaining the galaxy only if $C(S_{\rm p},S_{\rm int})\geq R$. In addition, we must account for the fact that the completeness function itself is very uncertain for $C(\sp,\sint)<0.5$ (e.g., Fig.~6 in \citealp{Zwaan2004}). As in \cite{Zwaan2010}, we therefore only retain galaxies with $C(S_{\rm p},S_{\rm int})\geq0.5$. These completeness cuts reduce the total number of simulated sources in the HIPASS volume to about $8\%$, while reducing the number of observed sources by 514, i.e.,~by an additional $9.8\%$ after the $1.3\%$ mass cut. This concludes the construction of the reference samples.

\begin{table}[t]
	\centering
	\normalsize
	\begin{tabular}{lcc}
	\hline\hline \\ [-1.5ex]
	Selection criterion 					& trims HIPASS		& trims \sax 	\\ [1.0ex]
	\hline \\ [-1.5ex]
	\multicolumn{3}{l}{\textit{Reference sample:}}							\\ [0.5ex]
	$\rm dec\leq+25^\circ$				& no				& yes		\\
	$cz<=12,700\rm~km\,s^{-1}$			& no				& yes		\\
	$\mha\geq10^8\,\hhmsun$			& yes			& yes		\\
	$\w\geq25\rm~km\,s^{-1}$			& no				& yes		\\
	$C(S_{\rm p},S_{\rm int})\geq0.5$		& yes			& yes		\\
	$C(S_{\rm p},S_{\rm int})\geq R$		& no				& yes		\\ [1.5ex]
	\hline \\ [-1.5ex]
	\multicolumn{3}{l}{\textit{Additional criteria for $\vc$-sample:}}				\\ [0.5ex]
	$\rm dec\leq+2^\circ$				& yes			& yes		\\
	$0.86\geq R$						& no				& yes		\\ 
	$i\geq45^\circ$						& yes			& yes		\\ [1.5ex]
	\hline
	\end{tabular}
	\caption{\upshape\raggedright The upper list shows the criteria applied to all data presented in this paper. The lower list shows the additional selection criteria applied when comparing circular velocities $\vc$, which require optical inclination measurements. Here, $R\in[0,1]$ denotes a random number drawn separately for every galaxy and equation.}
	\label{tab_selection_criteria}
\end{table}

\subsubsection{$\vc$-sample}

To compare observed and simulated values of $\vc$, the reference sample must be further reduced to a HOPCAT equivalent sample, i.e., a subsample with optically measured inclinations.

\textit{Volume truncation:} We must exclude the galaxies with $\rm dec>+2^\circ$, for which optical inclinations are not readily available in HIPASS. In doing so, the number of observed and modeled objects shrinks by roughly $20\%$.

\textit{HOPCAT completeness:} Out of all galaxies in the reference sample with $\rm dec\leq+2^\circ$ only 86\% yield optical inclinations. Most of the remaining objects lie too close to the galactic plane, where the stellar foreground deteriorates extragalactic optical imaging. To account for this incompleteness, we reduce the number of simulated galaxies to 86\% by only retaining the objects satisfying $0.86\leq R$, where $R\in[0,1]$ is a random number.

\textit{Inclination selection:} Galaxies with inclinations close to face-on exhibit poor inclination measurements, which, given their small values of $i$, result in highly uncertain values of $\vc$ when using \eq{eq_vc}. As in \cite{Zwaan2010} we therefore only retain galaxies with $i\geq45^\circ$, hence reducing the sample sizes by an additional $29\%$.

\section{Comparison between HIPASS and S$^3$-SAX}\label{section_comparison}

Given the identically selected samples of observed and simulated galaxies, we shall now compare the statistics of the \ha\ line profiles. This comparison will be carried out both at the level of direct source counts (Section \ref{subsection_number_counts}) and at the level of space density functions (Section \ref{subsection_space_densities}).

\subsection{Sample size}

Let us first consider the raw size of the observed and simulated samples given in Tab.~\ref{tab_sample_sizes}.

The mean number of sources in the five simulated reference samples is about 4926 with a standard deviation of 482. This standard deviation is significantly higher than the Poisson shot noise of $\sim\sqrt{4926}\approx70$, demonstrating the non-negligible effect of large scale structure in HIPASS. The number of observed sources in the reference sample is clearly consistent with the simulation. We therefore expect the normalization of corresponding source count statistics and space density functions to be consistent between observation and simulation.

By contrast, the mean number of sources in the simulated $\vc$-samples (about 2013) undershoots the number of observed sources by about 339 or 14\%. This difference is slightly larger than the characteristic value of cosmic variance of 267, estimated from the standard deviation of the object-numbers in the five simulated $\vc$-samples. As argued in Section \ref{subsection_number_counts}, this moderately significant difference between the sizes of the observed and simulated $\vc$-samples is at least partially explainable by a small fraction of inaccurate inclination measurements in HOPCAT. Those tend to assign high inclinations ($i\geq45^\circ$) to objects, which in actual fact have low inclinations ($i<45^\circ$) and should hence be removed from the observed $\vc$-sample.

\subsection{Source counts}\label{subsection_number_counts}

A refined statistical analysis consists of counting the number of galaxies, binned by specific galaxy properties. The properties of particular interest are the \ha\ linewidth $\w$ and the circular velocity $\vc$ defined by \eq{eq_vc}. For completeness we also analyze the statistics of the \ha\ mass $\mha$. The source counts of $\mha$ and $\w$ are derived from the reference samples. In turn, the source counts of $\vc$, which require inclination measurements, must be performed using the smaller $\vc$-samples.

\begin{figure*}
	\includegraphics[width=\textwidth]{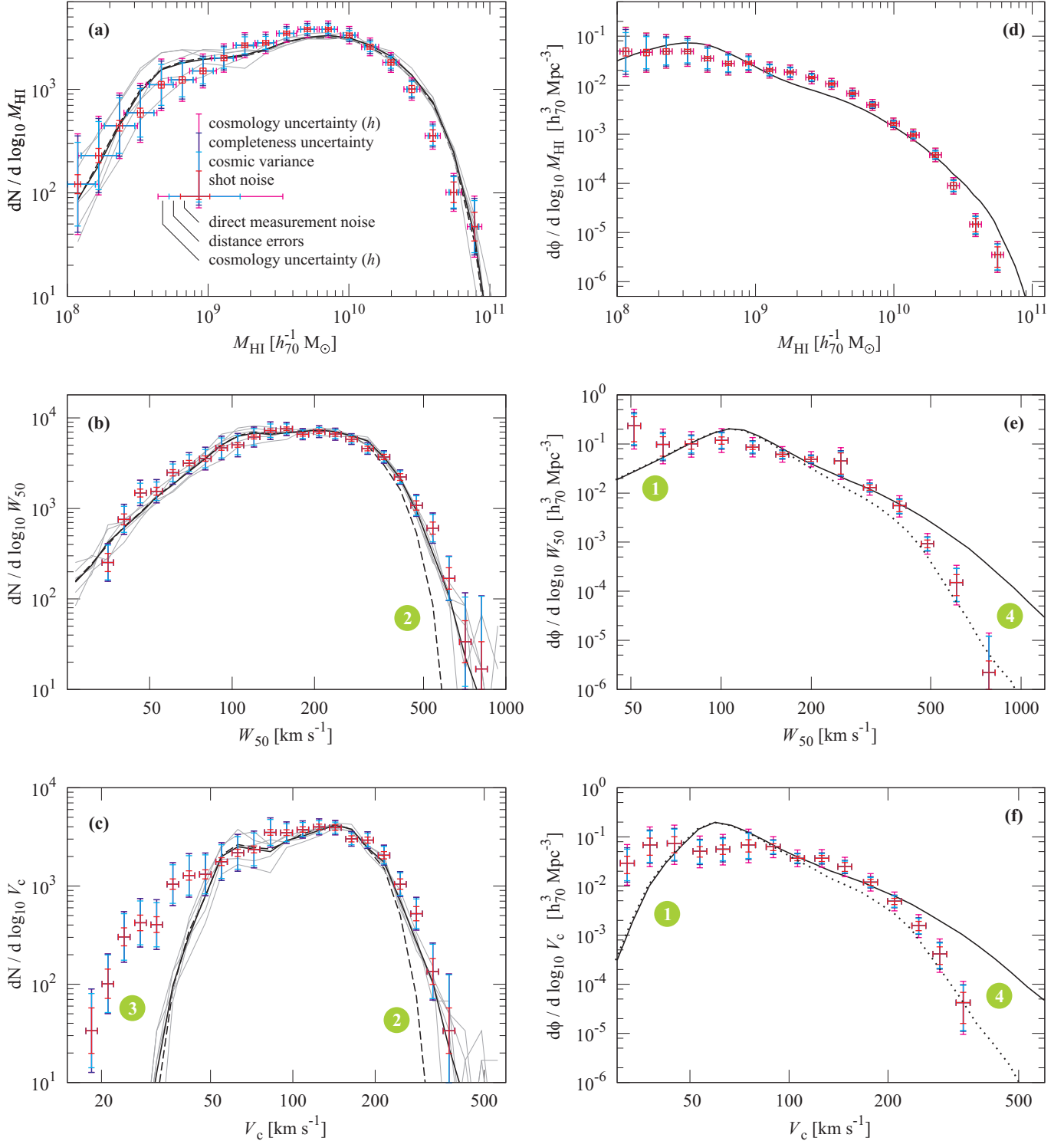}
	\caption{(Color online) Statistical comparison between HIPASS and \sax. Error bars represent various measurement uncertainties described in Tab.~\ref{tab_errors}. Thin grey solid lines represent the five statistically independent simulations, while the thick solid lines represent their geometric averages. Dashed lines delineate the same averages if source-confusion is not accounted for. Dotted lines represent only late-type galaxies, excluding S0 and E-types. Green numbers denote artifacts discussed in Section \ref{section_comparison}; they match the numbers in the abstract. (a) counts of masses $\mha$ in the reference samples; (b) counts of linewidths $\w$ in the reference samples; (c) counts of circular velocities $\vc\equiv\w/(2\sin i)$ in the $\vc$-samples; (d) \ha~MF as derived from HICAT by \cite{Zwaan2005} and predicted using all model-galaxies in the Millennium box; (e) space density function of $\w$ for galaxies of all Hubble-types as derived from HICAT by \cite{Zwaan2010} and predicted using all model-galaxies; (f) \ha~VF for galaxies of all Hubble-types with inclinations $i\geq45^\circ$ as derived from HICAT and HOPCAT by \cite{Zwaan2010} and predicted using all model-galaxies. Since the observed data points use all Hubble-types, the data in panels (e) and (f) must be compared against the solid function. The dotted function (only late-types) nonetheless provides a better fit, because the predicted class of gas-poor early-types was simply not detectable by HIPASS (details in section \ref{subsection_space_densities}).}
	\label{fig_overview}
\end{figure*}

Figs.~\ref{fig_overview}a--c show the observed (bars) and simulated (lines) counts of $\mha$, $\w$, and $\vc$, respectively. The grey solid lines correspond to the five individual simulations, while the black lines represent the geometric means of these functions. Variations between the five models are due to cosmic variance. The observed source counts exhibit several error bars, representing the uncertainties described in Tab.~\ref{tab_errors}. Some of these uncertainties are statistical, while others are systematic and thus correlated across different bins.

\begin{figure}
	\includegraphics[width=\columnwidth]{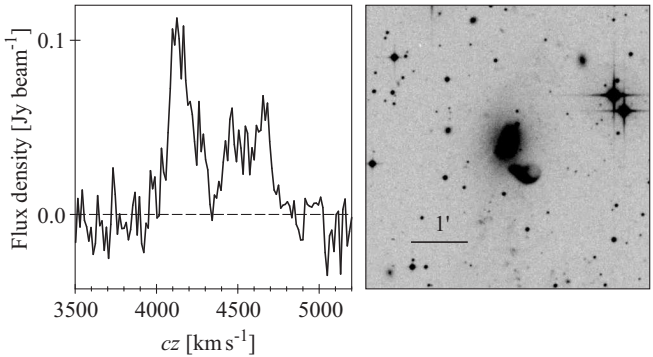}
	\caption{Example of a typical confused source in HICAT. This object (HIPASSJ1347-30) was assigned a single width of $\w=653.6~\kms$ (without confusion flag), the highest value of any source with $\sint>50~\rm{Jy}~\kms$. The SuperCOSMOS optical $b_J$-band image suggests that the \ha\ emission line is a combination of two merging systems.}
	\label{fig_confusion_real}
\end{figure}

\begin{table}[t]
	\centering
	\normalsize
	\begin{tabular*}{\columnwidth}{lcc}
	\hline\hline \\ [-1.5ex]
					& ~~reference sample~~	& ~~$\vc$-sample~~	\\ [1.0ex]
	\hline \\ [-1.5ex]
	Observation		& 4732			& 2352		\\
	Simulation 1		& 4699			& 1839		\\
	Simulation 2		& 5268			& 2194		\\
	Simulation 3		& 5034			& 2103		\\
	Simulation 4		& 4212			& 1642		\\
	Simulation 5		& 5416			& 2285		\\ [1.5ex]
	\hline
	\end{tabular*}
	\caption{\upshape\raggedright Number of sources in each sample. The selection criteria are listed in Tab.~\ref{tab_selection_criteria} and the simulated volumes are shown in Fig.~\ref{fig_millennium}.}
	\label{tab_sample_sizes}
\end{table}

The observed and simulated $\mha$ counts in Fig.~\ref{fig_overview}a are moderately consistent. Four of the five models and the mean model show a slight bump around $\mha\approx4\cdot10^8\msun$. This seems to be a feature of the particular SAM chosen here, since it is also present in the $b_J$-band LF of the same SAM (see Fig.~8 right of \citealp{Croton2006}), but absent in other SAMs building on the Millennium simulation \citep[e.g.][]{Baugh2005}.

Fig.~\ref{fig_overview}b is the central plot of this paper and extends on Fig.~\ref{fig_essential}. It demonstrates that the simulated linewidths $\w$ are fully consistent with the observed ones. We emphasize that this consistency requires that the simulated and observed samples are constructed according to identical selection criteria (see Tab.~\ref{tab_selection_criteria}). Experimenting with different completeness functions $C$ further revealed the importance of using the smooth completeness function $C(\sp,\sint)$ provided for HICAT and NHICAT. A hard sensitivity limit, i.e., $C(\sp,\sint)$ as a step-function, is not sufficient in that it induces variations larger than the error bars. Moreover, accounting for the confusion of sources turns out to be vital. If instead all individual galaxies in the simulated sky were considered distinguishable, then the mean source counts are given by the dashed line in Fig.~\ref{fig_overview}b. The difference is most pronounced at the largest linewidths of $\w\gtrsim500~\kms$ (artifact `2'). Thus, the largest values of $\w$ in the observed data are mostly due to confused sources, i.e., galaxies within the same telescope beam and with \ha~line profiles overlapping in frequency space. In constructing the original HICAT dataset \citep{Meyer2004} an effort was made to flag and separate sources exhibiting confused \ha\ line profiles. About 9\% of the sources with $\w\geq500~\kms$ in the reference-sample have been flagged as confused (as opposed to 7\% in the whole reference sample). By contrast, our modelling revealed that most sources with $\w\geq500\kms$ are confused. This means that it may be impossible to identify most instances of confusion by relying exclusively on the information in the HIPASS data. An example of a confused source is shown in Fig.~\ref{fig_confusion_real}.

\begin{figure}
	\includegraphics[width=\columnwidth]{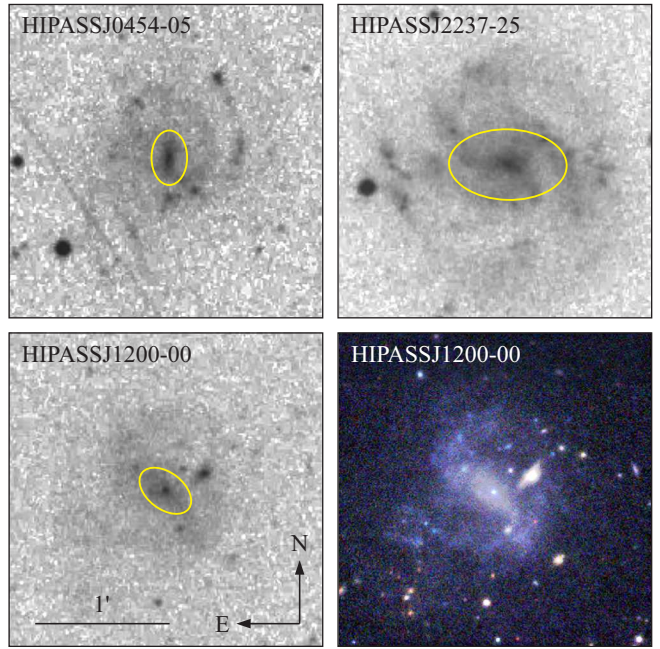}
	\caption{(Color online) Three examples of the few galaxies in HOPCAT (about 6\% of all HICAT/HOPCAT objects) with uncertain/inaccurate shape parameterizations. The three greyscale images are the $b_J$-band maps from the SuperCOSMOS Sky Survey \citep{Hambly2001} used in HOPCAT, while the false-color image ($i$-band in red, $r$-band in green, $g$-band in blue) shows a corresponding deep image obtained by the Galaxy And Mass Assembly (GAMA) survey \citep{Driver2011}. Yellow ellipses represent the fits quoted in HOPCAT; they all overestimate the inclinations of the galaxies. [Note that position angles in HOPCAT are given anti-clockwise from west rather than north.]}
	\label{fig_hopcat}
\end{figure}

The counts of circular velocities $\vc$ are shown in Fig.~\ref{fig_overview}c. The models are consistent with the observations for $\vc>50~\kms$, but drastically differ for smaller velocities (artifact `3'). The only major difference between Fig.~\ref{fig_overview}b and Fig.~\ref{fig_overview}c is the inclination-correction [see eq.~(\ref{eq_vc})]; therefore the excess of observed sources with $\vc<50~\kms$ suggests an issue with their inclinations. A systematic visual inspection of the $b_J$-band images of the SuperCOSMOS Sky Survey used in HOPCAT uncovered that a vast majority ($>90\%$) of the galaxies with $\vc<50~\kms$ (about 11\% of the 2352 objects in the $\vc$-sample or 6\% of all 4315 galaxies in HICAT/HOPCAT) are problematic. They are either too faint or too irregular for an optical estimation of the inclination, or they simply exhibit erroneous shape parameterisations. Fig.~\ref{fig_hopcat} displays three representative examples of the latter case. The ellipses in Fig.~\ref{fig_hopcat} represent the original parameterization in terms of minor axis, major axis, and position angle. The axis ratios of these ellipses imply inclinations $i>45^\circ$ [via eq.~(\ref{eq_i})]. To the naked eye, however, these three galaxies are nearly face-on spiral disks ($i<45^\circ$), especially in the multi-color image of the source HIPASSJ1200-00, which is about two magnitudes deeper than SuperCOSMOS. Using the `correct' inclination for this source rather than that suggested by HOPCAT, increases $\vc$ roughly by a factor two. Since the correct inclination is then below $45^\circ$, this source would be rejected from the $\vc$-sample and thus disappear from Fig.~\ref{fig_overview}c. In conclusion, there is a small fraction of incorrect shape identifications in HOPCAT, which happens to dominate the low-end of the $\vc$ counts. Incidentally, this also explains the asymmetric scatter skewed towards low rotational velocities in the HOPCAT-based TFR (upper panels in Fig.~3 in \citealp{Meyer2008}).

%\begin{figure}
%	\includegraphics[width=\columnwidth]{fig_inclinations}
%	\caption{(Color online) $\cos(i)$-distribution in the observed (rectangles) and simulated (lines) $\vc$-samples. Thin solid lines correspond to the five independent realizations of the simulation, while the thick solid line denotes their arithmetic average.}
%	\label{fig_inclinations}
%\end{figure}

\begin{table*}[t]
	\centering
	\normalsize
	\begin{tabular*}{\textwidth}{llp{11.1cm}}
	\hline\hline \\ [-1.5ex]
	Error name & Type & Explanation \\ [1.0ex]
	\hline \\ [-1.5ex]
		cosmology uncertainty ($h$)~ & systematic~ & Variations of the Hubble parameter $h$ on the interval $[0.65,0.75]$. This interval contains the fiducial value of $h=0.704^{+0.013}_{-0.014}$ \citep[7-year WMAP+BAO,][]{Jarosik2011}, as well as the recently found $h=0.743\pm0.021$ \citep[Spitzer and Hubble space telescope,][]{Freedman2012}. Note that it is important to plot $h$-related uncertainties, since observed and simulated data scale differently with $h$; e.g., simulated masses are in units of $h^{-1}\msun$, while observed masses are in units of $h^{-2}\msun$. \\
		completeness uncertainty & systematic & Approximate $67\%$-confidence intervals associated with systematic uncertainties of the completeness function $C$ [see \eq{eq_completeness}] in the range $C>0.5$. We calculate these uncertainties as $\Delta C=0.4(1-C)/C$, which approximately matches the error bars in Figs.~2 and 6 of \cite{Zwaan2004}. Note that variations in the completeness function would alter the simulated data rather than the observed data. In Fig.~\ref{fig_overview} these error bars have only been plotted on the observed data for graphical convenience. \\
		cosmic variance & statistical & $67\%$-confidence intervals associated with cosmic variance, as determined from the standard deviation between the five simulated samples. \\
		shot noise & statistical & Approximate $67\%$-confidence intervals associated with Poisson shot noise, calculated as the square root of the number of sources in the bin. \\
		direct measurement noise & statistical & Approximate $67\%$-confidence intervals associated with telescope noise and limited frequency resolution (Section \ref{subsection_observation}). \\
		distance errors & statistical & $67\%$-confidence intervals associated with errors in the spectroscopic distance measurement, assuming average line-of-sight peculiar velocities of $300~\kms$. This uncertainty is largest from nearby sources and therefore largest for low-mass galaxies in HIPASS. \\ [1.5ex]
	\hline
	\end{tabular*}
	\caption{\upshape\raggedright Explanation of the different error bars shown in Fig.~\ref{fig_overview}. The error `type' refers to whole sample. For example, distance errors due to peculiar velocities are systematic for an individual source, but statistical at the level of a sample of sources with random peculiar motions. In combining multiple errors into a single error bar, statistical errors are added in quadrature, while systematic errors are added linearly.}
	\label{tab_errors}
\end{table*}

\subsection{Space densities}\label{subsection_space_densities}

The source counts presented in the previous section depend on the selection criteria of the survey listed in Tab.~\ref{tab_selection_criteria}. Survey-independent and thus more fundamental statistical measures are the space density functions $\phi\equiv\d N/\d V$. These functions represent the absolute number of sources, detected or not, per unit of cosmic volume and per unit of galaxy properties, such as $\mha$ (\ha~MF) or $\vc$ (VF). Evaluating these functions from empirical data requires inverting the completeness function, as well as removing the effects of cosmic variance. This is achieved by the two-dimensional stepwise maximum likelihood (2DSWML) method developed by \cite{Zwaan2003} and applied by \cite{Zwaan2005} and \cite{Zwaan2010} to recover the observed space density functions of $\mha$, $\w$, and $\vc$, shown in Figs.~\ref{fig_overview}d--f. Note that the data shown here include all Hubble-types. Figs.~\ref{fig_overview}d--f also display the simulated counterparts (solid lines), obtained simply by binning all galaxies contained in the redshift $z=0$ box of the Millennium simulation. This box is large enough for cosmic variance to be neglected. However, the observed space density functions still obey the same cosmic variance as the respective source counts. Therefore the cosmic variance uncertainty is plotted with the observed data, although we derive its value from the variations between the five simulated source counts.

Fig.~\ref{fig_overview}d reveals that the simulated and observed \ha~MFs are only marginally consistent in the sense that the simulation falls within the error bars for about $50\%$ of the data points rather than $67\%$. The fact that the agreement was slightly better in source count statistics of Fig.~\ref{fig_overview}a might indicate a minor artifact in the reconstruction of the observed \ha~MF. For example, as suggested by \cite{Zwaan2004}, the `true' completeness function $C$ exhibits a slight dependence on the shape of the \ha\ line profile (single-peaked, double-peaked, flat-top) in addition to the main dependence on $\sp$ and $\sint$. This small higher-order effect could be captured by extending the 2DSWML method to 3D using $C(\sp,\sint,shape)$.

Figs.~\ref{fig_overview}e and \ref{fig_overview}f suggest clear inconsistencies between the models and observations. In the small velocity range, these inconsistencies (artifact `1') directly relate to the mass resolution limit of the Millennium simulation. This limit implies a significant incompleteness of simulated objects with $\w\lesssim80~\kms$ and $\vc\lesssim50~\kms$ (and $\mha<10^8\msun$ to the left of Fig.~\ref{fig_overview}d). In turn, this mass-limit is probably linked to the spurious bumps around $\w\approx120~\kms$ and $\vc\approx70~\kms$.

\begin{figure}[b]
	\includegraphics[width=\columnwidth]{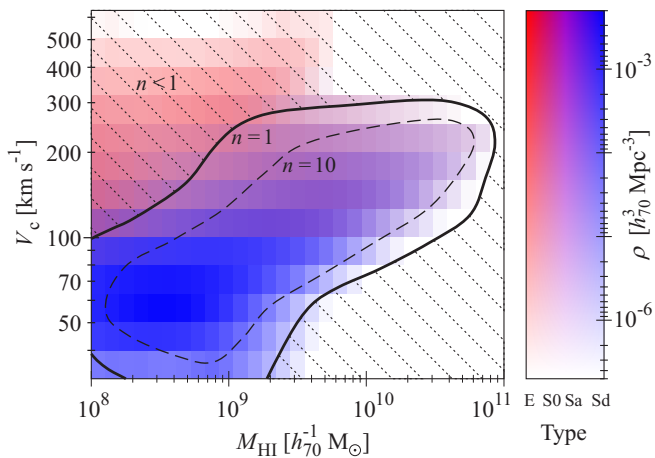}
	\caption{(Color online) Predicted space density $\rho(\mha,\vc)$ per pixel of size $\Delta\log_{10}\mha=\Delta\log_{10}\vc=0.1$, colored according to the average galaxy type in each pixel. Curved lines represent isolines of the number $n$ of expected detections per pixel; the diagonal shading ($n<1$) highlights the blind zone of HIPASS.}
	\label{fig_ellipticals}
\end{figure}

A more subtle feature in Figs.~\ref{fig_overview}e and \ref{fig_overview}f are the significant deviations at $\w\gtrsim500~\kms$ and $\vc\gtrsim200~\kms$ (artifact `4'). Those deviations are absent in the corresponding source counts of Figs.~\ref{fig_overview}b and \ref{fig_overview}c. A systematic investigation of the simulated galaxies in this high-velocity regime reveals them to be dominated by early-type galaxies of numerical Hubble-type $T\leq0$ (E, S0) hosting low-mass, but fast-rotating \ha\ disks. Excluding those objects from the simulation modifies the predicted functions in Figs.~\ref{fig_overview}e and \ref{fig_overview}f to the dot-dashed lines, which are in much better agreement with the observed data, as already noted by \cite{Zwaan2010}. In other words, the model predicts that the high-end of the VF is dominated by gas-poor early-type galaxies, but it also predicts that HIPASS is unlikely to detect these galaxies; hence the consistent source counts. To show this explicitly, let us calculate the maximal comoving distance $D_{\rm max}$ (in Mpc) out to which a galaxy $\{\mha,\vc\}$ can be detected in the sense that the completeness function $C$ drops to $50\%$ at that distance. Substituting $\sp$ for $10^3\sint\vc^{-1}$ (approximation for $i>45^\circ$) and $\sint$ for $4.2\cdot10^{-6}\mha D_{\rm max}^{-2}$ (approximation for $z\ll1$), $C(\sp,\sint)=0.5$ [using eq.~(\ref{eq_completeness})] numerically solves to $D_{\rm max}^2\approx7\cdot10^{-6}\mha\,\exp(-0.4\vc^{0.34})$ for HICAT and NHICAT. The cosmic volume $V_{\rm max}$ (in Mpc$^3$), in which HIPASS can detect a galaxy specified by $\{\mha,\vc\}$ then becomes $V_{\rm max}\approx0.63\cdot(4\pi/3)D_{\rm max}^3$ , where 0.63 is the sky-coverage of HIPASS, i.e.,
\be
	V_{\rm max}(\mha,\vc) \approx 5\cdot10^{-8}\mha^{3/2}\exp\left(-0.6\vc^{0.34}\right).
\ee
On the other hand, the \sax~model allows us to predict the space-density $\rho(\mha,\vc)$ of a source $\{\mha,\vc\}$, defined as the average number of sources per Mpc$^3$ within a pixel $\{\log_{10}\mha\pm\Delta/2,\log_{10}\vc\pm\Delta/2\}$ (here using $\Delta=0.1$). The product
\be
	n(\mha,\vc) \equiv V_{\rm max}(\mha,\vc)\rho(\mha,\vc)
\ee
then approximates the predicted number of HIPASS detections per pixel in the $\{\mha,\vc\}$-plane. Fig.~\ref{fig_ellipticals} displays $\rho(\mha,\vc)$ colored by galaxy type with isolines of $n(\mha,\vc)$. The region $n(\mha,\vc)<1$ contains less than one detection per pixel and thus represents a `blind zone' of HIPASS. This blind zone contains the gas-poor ($\mha\lesssim10^9~\msun$), fast-rotating ($\vc\gtrsim200~\kms$) early-type galaxies predicted by the model. Since HIPASS is very insensitive to these galaxies, it is simply unable to recover the predicted high-end of the VF. Surveys deeper than HIPASS are needed to verify whether the predicted amount of massive gas-poor early-type galaxies  is correct. For now, it seems safe to conclude that the HIPASS VF approximates the VF of late-types, even if no Hubble-type cut is applied to the dataset. On a side-note, the deeper ALFALFA survey does indeed find significant differences in the high-velocity end of the velocity function (e.g.~Fig.~4 in \citealp{Papastergis2011}).

In principle, the artifacts `2' and `3' of Figs.~\ref{fig_overview}b and \ref{fig_overview}c are still present in Figs.~\ref{fig_overview}e and \ref{fig_overview}f, but they are occluded by the even stronger artifacts `1' and `4'. This shows that the comparison between models and observations is less prone to spurious artifacts, when performed using source counts. Furthermore, within the source counts, $\w$ is a less problematic quantity than $\vc$ due to artifact `3'.

\section{Discussion}\label{section_discussion}

This section discusses the physical implications of the excellent consistency between observed and simulated \ha\ linewidths, as well as potential applications.

\begin{figure}[b]
	\includegraphics[width=\columnwidth]{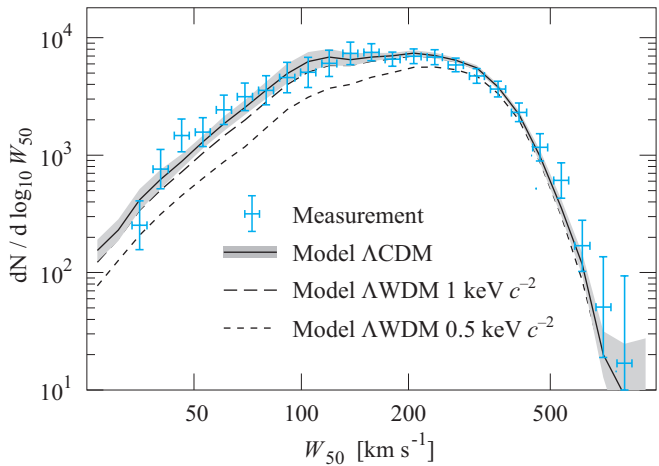}
	\caption{(Color online) Analogous plot to Fig.~\ref{fig_essential}, but with additional lines for two alternative dark matter models assuming finite particles masses of $1\kevc$ and $0.5\kevc$. The error bars sum up all the statistical and systematic uncertainties considered in this work (see Fig.~\ref{fig_overview}) and the grey shading denotes the standard deviation of five independent simulated reference samples for CDM.}
	\label{fig_wdm}
\end{figure}

\subsection{Interpretation of the consistency of $\w$}\label{subsection_interpretation}

What does the consistency between the observed and modeled $\w$-counts (Fig.~\ref{fig_overview}b) tell us? Does it strengthen the case of the $\Lambda$CDM model or does it merely manifest the empirical tuning of the free parameters in the galaxy-model? There is, as argued here, a bit of both. The local galaxy stellar MF in the model has been adjusted indirectly by tuning the feedback from star formation and black holes on the interstellar medium to reproduce the observed $b_J$-band and $K$-band LFs \citep{Croton2006}. Moreover, the radii of galaxies match the locally observed mean stellar mass-to-scale radius relation \citep{Obreschkow2009b}. One might therefore expect the galaxy rotations, which depend roughly on mass and radius, to align with local observations. In this argument, it should nonetheless be emphasized that the free model-parameters (feedback coefficients and the spin ratio of baryonic matter to dark matter) have only been varied within the restricted ranges consistent with current high-resolution observations and high-resolution simulations. Therefore, we can at least conclude that the consistency of the $\w$-counts in Fig.~\ref{fig_overview}b confirms $\Lambda$CDM within the current uncertainties of galaxy-modelling.

Moreover, it is worth emphasizing that the relation between stellar mass and scale radius is subject to very large scatter, both observationally and in the model \citep[Fig.~2 in][]{Obreschkow2009b}. Therefore, even if the mean relation between stellar mass and scale radius is fixed to observations, this merely corresponds to an overall normalization of the VF and the corresponding $\w$-counts. The details of these functions depend on the shape of the multi-dimensional probability-distribution of halo mass, stellar mass and disk scale radius. This shape has not been constrained by empirical fits. Instead, it depends directly on the masses, spins, and merging histories of the dark halos in the Millennium simulation. This argument increases the support of $\Lambda$CDM.

\subsection{Constraints on the dark matter type}\label{subsection_wdm}

Quantifying the degree to which the $\w$-counts support $\Lambda$CDM is of course a more delicate affair. For example, what is the actual range of allowed dark matter particle masses $\mdm$, assumed infinite in CDM but finite in Warm Dark Matter (WDM) models? Answering this question would require a large array of different WDM models, similar to the Millennium simulation, equipped with SAMs, where all the uncertainties associated with every free parameter are tackled down to the $\w$-counts. The mammoth numerical requirements of this task lie at the edge of current super-computing capacities.

Here, we limit the analysis to a first order approximation of the variation of the $\w$-counts as function of $\mdm$, keeping the free parameters of the galaxy-model fixed to their best values in $\Lambda$CDM. This approximation is obtained by rescaling the number density of each galaxy in \sax, one-by-one, by $\phi_{\rm WDM}(M_{\rm halo})/\phi_{\rm CDM}(M_{\rm halo})$, where $M_{\rm halo}$ is the mass of the halo containing the galaxy and $\phi_{\rm WDM}(M_{\rm halo})$ and $\phi_{\rm CDM}(M_{\rm halo})$ are the local halo MFs of WDM and CDM halos, respectively. These MFs are modeled analytically by evolving the initial density field using the formulation of \cite{Sheth1999}. WDM models are obtained by subjecting the initial CDM power spectrum to a transfer function following \citet{Bode2001}. For consistency, these calculations were performed using the cosmological parameters of the Millennium simulation. 

Fig.~\ref{fig_wdm} shows the $\w$-counts for CDM and two WDM scenarios with particle masses $\mdm=1\kevc$ and $\mdm=0.5\kevc$, respectively. Although the observed $\w$-counts are only marginally consistent with $\mdm=1\kevc$ and inconsistent with $\mdm=0.5\kevc$, those WDM cosmologies need \textit{not} to be incompatible with the observed $\w$-counts. In fact, we cannot exclude that varying the free parameters of the SAM within the currently allowed ranges can bring the WDM models in line with the observed data. However, Fig.~\ref{fig_wdm} conveys that if all free parameters in the galaxy-model can be replaced by independently determined precise values, then the $\w$-counts from HIPASS can indeed discriminate between CDM and WDM with $1\kevc$ particles. 

\begin{figure}[t]
	\includegraphics[width=\columnwidth]{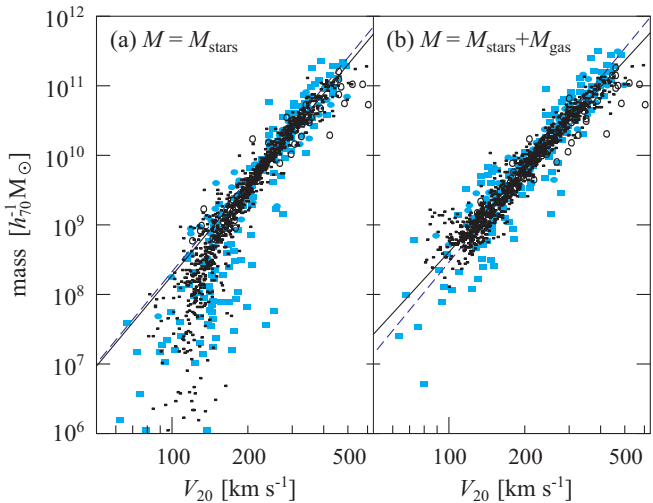}
	\caption{(Color online) Relationships between edge-on \ha~linewidths $V_{20}$ and two mass tracers for local galaxies. A representative sample of $10^3$ simulated galaxies is shown as black points (spiral galaxies) and open circles (elliptical galaxies). Solid lines are uniformly weighted power-law fits to the simulated spiral galaxies; in Fig.~\ref{fig_tfr}a this fit uses only galaxies with $\ms>10^9\hhmsun$. The blue dots and dashed lines are observational data and power-law fits from \cite{McGaugh2000} and references therein.}
	\label{fig_tfr}
\end{figure}

\subsection{Tully-Fisher relation}

So far, we have shown that the \ha\ masses and circular velocities of the galaxies in the SAM (as modeled via \sax) are consistent with observations; and \cite{Croton2006} showed that the stellar masses are consistent with local observations as well. However, the fact that circular velocities and masses are independently consistent with observations does not, in fact, imply that their two-dimensional distribution is correct, too. Therefore, we shall finally discuss the two-dimensional distribution of circular velocities and baryon masses, i.e., the baryonic TFR. To remain consistent with observational standards the circular velocity is here approximated as $V_{20}$, defined as half the apparent \ha\ linewidth $W_{20}$ (measured at the 20\% peak flux level), corrected for inclinations. The observational data is drawn from \cite{McGaugh2000} and corrected for $h=0.73$. These data include galaxy types from dwarfs to giant spirals, whose values of $V_{20}$ have been recovered from \ha\ line measurements, corrected for inclinations drawn from optical imaging. Only inclinations above $45^\circ$ were retained to restrict the uncertainties of $\sin i$. The comparison of these data against \sax\ in Fig.~\ref{fig_tfr} reveals a good consistency, although the observational scatter is 50\% larger than that of \sax. This difference is explainable by measurement uncertainties, especially regarding the inclination corrections in the low-mass end of Fig.~\ref{fig_tfr}a according to \citeauthor{McGaugh2000}. Additionally, the \sax-model probably underestimates the scatter in $V_{20}$ by ignoring the detailed substructure of \ha, such as turbulent mixing in mergers, high-velocity clouds, warps, and gas-rich satellites.

Unlike the baryonic TFR (Fig.~\ref{fig_tfr}b), the stellar mass TFR (Fig.~\ref{fig_tfr}a) clearly departs from a power-law relation for galaxies with $V_{20}<200~\kms$. As emphasized before (e.g.~\citeauthor{McGaugh2000}), this reflects the trend for high gas-fractions in low-mass galaxies and confirms that the TFR is fundamentally a relation between circular velocity and total mass, which is a function of the baryon mass \citep{Papastergis2012}.\vspace{1mm}

\section{Conclusion}\label{section_conclusion}

This paper presented a detailed comparison between the \ha\ lines is HIPASS and those in \sax, a cosmological model of galaxies with resolved \ha\ lines. The results can be condensed into a list of key messages.
\begin{enumerate}
	\item The \ha~linewidths of the \sax-model are consistent with those measured from HIPASS (Fig.~\ref{fig_overview}b). Hence, \textit{observed \ha~linewidths are consistent with \lcdm\ at the resolution of the Millennium simulation ($\mha\gtrsim10^8~\msun$, $\vc\gtrsim50~\kms$) within current galaxy formation models}. This does not contradict a possible breakdown of \lcdm\ at smaller masses \citep[e.g.][]{Zavala2009}.
	\item Galaxies with $\vc<50~\kms$ tend to be optically faint or irregular, thus suffering from large inclination uncertainties. To use these objects for physical applications, it is better compare simulations against apparent widths $\w$ rather than the inclination-corrected $\vc$ values.
	\item The model predicts that gas-poor early-type galaxies dominate the high-end of the VF. Yet the model also predicts that HIPASS is very insensitive to these galaxies because of their small $\mha$, large $\w$ (hence higher noise), and low space-density. To test whether gas-poor early-type galaxies really dominate the high-end of the VF deeper surveys are needed, but is seems safe to conclude that the HIPASS VF obtained using all observed galaxy types remains a VF of late-type galaxies.
	\item Most sources with $\w>500~\kms$ in HIPASS are found to be confused; hence confusion must be corrected in the high-end of the VF. This finding also applies to ALFALFA, because the $\sim4$ times higher spatial resolution of the Arecibo beam is nearly compensated by the mean redshift being $\sim3$ times higher.
	\item In general, \textit{$\w$ counts are the most reliable statistics of galaxy rotations}, since they can explicitly account for source confusion and complex completeness functions, and since they are not affected by inclinations. On the downside, $\w$ counts are less sensitive to cosmological parameters than velocity functions, since each value of $\w$ mixes galaxies of different masses seen at different inclinations. However, the $\w$ counts of HIPASS are nonetheless sensitive to the temperature of dark matter.
	\item In fact, if all free parameters in SAMs can be eliminated or at least constrained independently, the $\w$-counts derived from HIPASS can verify CDM against WDM with $1\kevc$ particles.
\end{enumerate}

These cosmological tests and prospects promise to become particularly fruitful when applied to future \ha\ surveys, such as the full ALFALFA survey and ultimately the ASKAP HI All-Sky Survey (WALLABY) with the Australian Square Kilometer Array Pathfinder (ASKAP). Those future surveys should be paralleled by equally sophisticated simulated counterparts, namely mock-skies produced from galaxy-models extending to considerably smaller masses and circular velocities than those based on the Millennium simulation. \\

D.~O.~acknowledges Elaine Sadler for her idea to model confused sources, as well as Simon Driver and Aaron Robotham for their assistance in preparing Fig.~\ref{fig_hopcat}. We thank the anonymous referee for a careful examination and very useful feedback.

%\bibliography{../Bibliography/astro}
%\bibliographystyle{../Bibliography/my_mn2e}

\end{document}